\documentclass[conference]{IEEEtran}
\usepackage{amsmath,graphicx,subcaption}
\usepackage{footnote,hyperref} 

\AtBeginDocument{%
	\providecommand\BibTeX{{%
			\normalfont B\kern-0.5em{\scshape i\kern-0.25em b}\kern-0.8em\TeX}}}

\makeatletter
\renewcommand\paragraph{\@startsection{paragraph}{4}{\z@}%
	{1.5ex plus .2ex minus .3ex}%
	{-0.5em}%
	{\normalsize\bf}}
\makeatother

\begin{document}
	%
	\title{CROMOSim: A Deep Learning-based Cross-modality Inertial Measurement Simulator}
	\author{
		\IEEEauthorblockN{Yujiao Hao}
		\IEEEauthorblockA{%
			Department of Computing and Software\\McMaster University\\
			Hamilton, Ontario,
			Canada\\
			Email: haoy21@mcmaster.ca}
		\and
		\IEEEauthorblockN{Boyu Wang}
		\IEEEauthorblockA{%
			Department of Computer Science\\Western University\\
			London, Ontario,
			Canada\\
			Email: bwang@csd.uwo.ca}
		\and
		\IEEEauthorblockN{Rong Zheng}
		\IEEEauthorblockA{%
			Department of Computing and Software\\McMaster University\\
			Hamilton, Ontario,
			Canada\\
			Email: rzheng@mcmaster.ca}
		

	}
	\maketitle
	
	\begin{abstract}
		With the prevalence of wearable devices, inertial measurement unit (IMU) data has been utilized in monitoring and assessment of human mobility such as human activity recognition (HAR). Training deep neural network (DNN) models for these tasks require a large amount of labeled data, which are hard to acquire in uncontrolled environments. To mitigate the data scarcity problem, we design CROMOSim, a cross-modality sensor simulator that simulates high fidelity virtual IMU sensor data from motion capture systems or monocular RGB cameras. It utilizes a skinned multi-person linear model (SMPL) for 3D body pose and shape representations, to enable simulation from arbitrary on-body positions. A DNN model is trained to learn the functional mapping from imperfect trajectory estimations in a 3D SMPL body tri-mesh due to measurement noise, calibration errors, occlusion and other modeling artifacts, to IMU data. We evaluate the fidelity of CROMOSim simulated data and its utility in data augmentation on various HAR datasets. Extensive experiment results show that the proposed model achieves a 6.7\% improvement over baseline methods in a HAR task.
	\end{abstract}
	
	%
	%

	\section{Introduction}
	Nowadays, inertial measurement units (IMUs) have become ubiquitously available in wearable and mobile devices. An important category of IMU-enabled applications is monitoring and assessment of human mobility, which aims to continuously track people's daily activities, analyze motion patterns and extract digital mobility bio-markers such as gait parameters in the wild. Increasingly, data-driven deep learning models have been developed for human activity recognition (HAR) ~\cite{wang2019deep,gu2021survey}. Despite their impressive performance, these models generally require a large amount of sensory data for model training. Unfortunately, it is challenging to collect high-quality IMU data in the wild while data collected from controlled settings where subjects are asked to perform certain activities often have very different characteristics from those in freestyle motions~\cite{vaizman2017recognizing}. 
	
	The scarcity of IMU data for HAR  is evident when compared with the richness of other data sources. PAMAP2~\cite{reiss2012introducing}, a popular dataset for HAR, includes 8 subjects with 59.67 minutes of samples per person. In contrast, AMASS~\cite{mahmood2019amass}, a motion capture  (MoCap) dataset, includes 2420.86 minutes of data and is growing; not to mention YouTube videos, which offer a practically infinite amount of action data. Therefore, to mitigate the ``small data'' problem, one possible solution is to convert data from other modalities to IMU, a process called {\it cross-modality simulation}. 
	
	Though several previous works explored the feasibility of simulating IMU sensor data from other data modalities (see Section \ref{section:RelatedWork}), a number of challenges remain. First, sensors are attached to human skin rather than directly to bone joints during data collection. Skeleton models are inadequate in representing human poses and shapes. Second, even with state-of-the-art (SOTA) solutions in computer vision, the extracted 3D human motion trajectories from monocular video clips are far from being perfect. Analytically computing IMU readings on such imperfect input sequences will result in large errors. However, if a deep learning model is adopted to learn the mapping between noisy motion trajectories and measured sensor readings, it is unclear how well such models generalize to arbitrary unseen on-body positions. 
	
	To tackle the above-mentioned challenges, we design and implement CROMOSim, a cross-modality IMU sensor simulator that simulates high fidelity virtual IMU sensor data from motion capture systems and monocular RGB cameras. It differs from existing work in several important aspects. First, it is based on 3D skinned multi-person linear (SMPL) models~\cite{loper2015smpl}. SMPL is capable of modeling muscle and soft tissue artifacts while the 2D or 3D skeleton representations adopted by other works are segment models without volumetric information. Second, we empirically demonstrated that the direct computation of IMU readings from motion trajectories extracted from videos is unreliable (in Section \ref{section:Evaluation}), even with filtering and interpolation techniques as the case of IMUSim~\cite{young2011imusim}. We instead design and train a neural network to learn the relationship between measured IMU readings and the noisy motion trajectories. Special cares have to be given to ensure the trajectories are represented in a global coordinate frame even if the videos are captured by moving cameras. Compared to existing IMU simulators, experiments show that CROMOSim achieves higher fidelity and superior performance in HAR tasks.  
	
	\section{Related Work}
	\label{section:RelatedWork}
	IMUSim \cite{young2011imusim} is amongst the first open-source tool to simulate IMU data from either MoCap data in the Biovision Hierarchy (BVH) format or a user-provided 3D position and orientation sequence. Given motion trajectories in a global frame, acceleration can be calculated by taking the second derivatives of positions over time. The resulting data has been used in existing work to pre-train human pose estimation (HPE) ~\cite{huang2018deep} and HAR models~\cite{xiao2020deep,takeda2018multi}. One drawback of this type of method is that none of these researches targets to simulate realistic IMU sensor readings, and gyroscope data is omitted.
	
	After that, simulating IMU readings from monocular RGB videos for data augmentation has attracted some attention in recent years. ZeroNet \cite{liu2021video} extracted finger motion data from videos and transformed them into acceleration and orientation information measured by IMU sensors. The authors of \cite{rey2019let} and its follow-up work~\cite{rey2020yet} simulated acceleration norms and/or angular velocity norms from human 2D poses for a HAR purpose. They differed from CROMOSim as both ZeroNet and Rey's sensor simulators simply avoided the video-based global motion tracking problem by limiting the human subjects‘ movement to a fixed camera scene (in-place motion), while we addressed such a problem in CROMOSim pipeline.
	Closest to our work is IMUTube~\cite{kwon2020imutube} and its extension in \cite{kwon2021approaching}, which aim at simulating full-body IMU data from moving camera videos captured in the wild. But confined by the skeleton body representation adopted, neither work can simulate realistic sensor readings from arbitrary on-body locations. Moreover, in IMUTube, the estimation of view depth and camera ego-motion is in two independent steps though the two are intrinsically coupled~\cite{kopf2021rcvd,mun2019unsupervised}. A wrongly predicted camera pose can lead to inaccurate view depth estimation and vice versa \cite{luo2020consistent}. In addition, the lifting of 2D postures to 3D poses module in IMUTube pipeline is more compute-intensive and error-prone, as it is a simple combination of existing technologies.
	
	\section{System Design}
	\label{method}
\begin{figure*}[t!]
			\centering
			\includegraphics[width=0.9\linewidth]{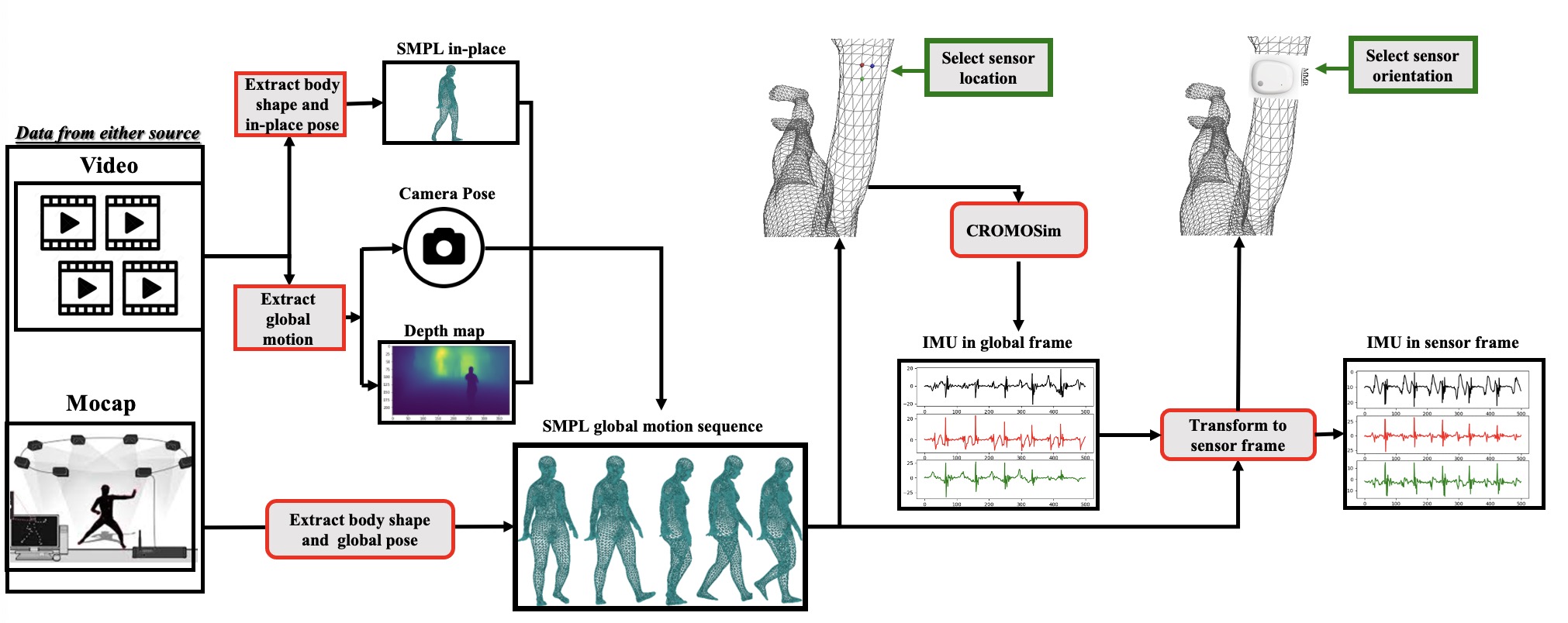}
			\caption{The CROMOSim pipeline. It takes either MoCap or monocular camera video data as input, converts them to SMPL represented global motion and body shape. The simulator then takes the SMPL model, specified sensor placement and orientation as input; predicts simulated IMU readings and transforms them back to the sensor coordinates frame.\label{fg:overall}}
	\end{figure*}

	CROMOSim is designed with several requirements in mind: i)  allowing arbitrary user-specified placement and orientation of target sensors,  ii) extensibility to different input data modalities and configurations, iii) flexibility to incorporate SOTA models to extract motion trajectories, and iv) high fidelity. To meet these requirements, the CROMOSim pipeline contains three function modules as shown in Fig.~\ref{fg:overall} : an input data processing module that extracts global human motion sequences from source data, a human body model that can fully represent the extracted sequences and can be sampled from any on-body location, and a simulator module that transforms noisy motion sequences to high-fidelity 3-axis accelerometer and gyroscope readings. Though the pipeline is extensible to other possible input data modalities such as millimeter wave radar and depth camera, we will focus on MoCap and monocular camera video here. A detailed illustration of each component will be provided in the following sections.
	
	
	\subsection{SMPL Model}
	An SMPL model represents 3D human body poses and shapes with a fine-grained full-body tri-mesh. Unlike skeleton or cylinder models that only capture joint poses, this parametric 3D representation provides a widely applicable and differentiable way to visualize a realistic 3D human body. There are three reasons to choose SMPL over other body models in CROMOSim. First, instead of measuring the movements of bones, IMU readings reflect the soft tissue dynamics at the location that a sensor is attached to. 
	Second, SMPL provides a pose and shape-dependent full-body tri-mesh that can be sampled at any on-body location. Third, since it is widely used in HPE research, many off-the-shelf models are available to extract SMPL representations from different data sources. 
	
\begin{figure}[h!]
		\centering
		\includegraphics[width=0.49\linewidth]{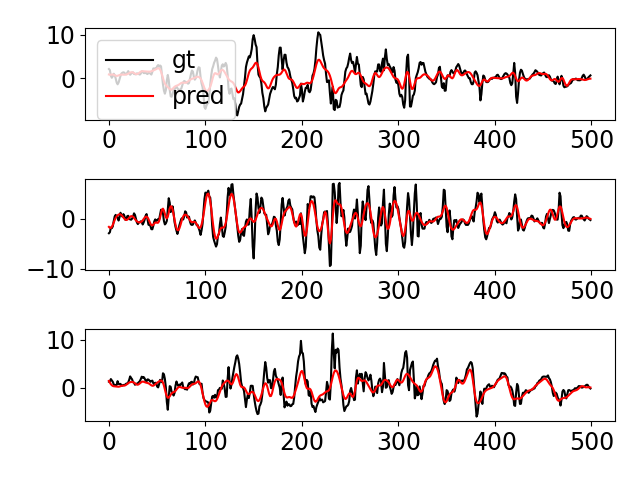}
		\includegraphics[width=0.49\linewidth]{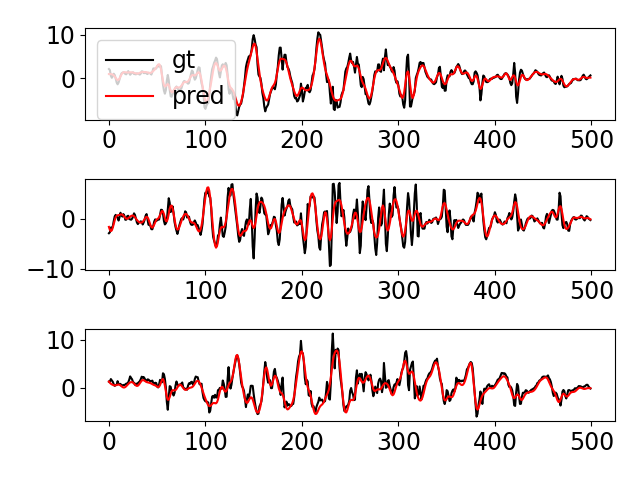}
		\caption{Comparison between analytically computed 3-axis accelerations from a skeleton representation and an SMPL model. Left: taking the sequence of pelvis joint positions as input, right: taking the sequence of SMPL lower back skin mesh positions as input.}
		\label{fg:smpl_skeleton}
	\end{figure}

	To see the difference between movements of joints in a skeleton model and SMPL skin mesh, we compare accelerations computed by taking second-order derivatives of the corresponding motion trajectories and ground-truth accelerometer readings over time. In Fig.~\ref{fg:smpl_skeleton}, red curves denote the calculated 3-axis accelerations while the black ones are accelerometer ground truth. Figures in the left column compare the accelerations at a pelvis joint in a skeleton model while figures in the right column compare those at an SMPL lower back skin mesh vertices. Clearly, the use of the SMPL skin mesh provides better agreements with the ground truth (e.g., in the interval [100,300]). Simulated data from the pelvic joint, on the other hand, fails to capture high-frequency acceleration components, which are most likely due to muscle and soft issue movements. This observation indicates that SMPL is a good candidate for an intermediate data representation of CROMOSim pipeline.
	
	\subsection{Input Data Processing}
	\subsubsection{From MoCap Data to SMPL Models}
	MoCap data consists of raw marker sequences collected by an optical motion capture system of high precision (usually with a position error $<1$ mm). MoSH++ \cite{mahmood2019amass} allows the fit of an SMPL model to MoCap data from a set of sparse markers.
	Prior to motion capture, a global coordinate system needs to be established during the calibration phase. As a result, the collected motion trajectories are expressed in the global frame. Under the assumption that the global frame is aligned with the inertial frame\footnote{Such an assumption is not restrictive as random rotation can be applied in further data augmentation to obtain data if the global and inertial frames differ.} , the SMPL mesh model can be used directly in subsequent processing. 
	
	\subsubsection{From Video Clips to SMPL Models}\label{sc:video}
	Extracting 3D human poses and shapes from monocular RGB videos is not trivial, especially when they are captured from moving cameras with unknown parameters, which is common in a locomotion-related video recorded in the wild. We propose to decompose such a problem into two sub-problems: a reconstruction of human global displacement and rotation; and an estimation of 3D in-place human motion and body shape. 
	
	\paragraph*{Estimating root joint global trajectory}
	A precise calculation of global displacement for the human subject is essential for a high-fidelity simulation of IMU data from RGB videos.  This requirement can be achieved by reconstructing the 3D motion trajectory of a fixed body position (a.k.a, the root joint), which can be inferred from the depth map of the human subject and camera parameters~\cite{prince2012computer}. 
	
	In CROMOSim, we adopt robust consistent video depth estimation (Robust CVD) method \cite{kopf2021rcvd}, a SOTA model to estimate consistent dense depth maps and camera poses from a monocular video. Robust CVD jointly estimates both outputs by solving an optimization problem over the entire video sequence. It is advantageous as the two outputs are intrinsically coupled and thus leads to higher accuracy (compared to the pipeline adopted by IMUTube). In the implementation, we locate the 2D torso joint positions in video frames using OpenPose~\cite{cao2019openpose}, and designate the pelvis as our root joint. In addition, depth reconstructed by robust CVD is reasonably accurate up to scale. To resolve scale ambiguity, an object of known size in the scene is needed. Prior knowledge regarding heights of subjects in the video, dimensions of fixtures (e.g., street lamps, road lanes) can be utilized. Subsequently, the predicted depth of the pelvis joint is re-scaled by an estimated scale factor. Since in some frames, the root joint is not visible or cannot be located well due to occlusion or poor lighting, we only extract root joint coordinates from the frames with high confident scores from OpenPose. Root joint coordinates in the remaining frames are then interpolated from the estimated ones, and a Kalman filter is applied to further smooth the resulting trajectory.
	
	\paragraph*{Body pose and shape estimation in camera frames}
	We adopt VIBE~\cite{kocabas2020vibe}, a SOTA method to directly estimate realistic 3D human poses and shapes from monocular videos. In the implementation, we make two extensions to VIBE. First, VIBE assumes a fixed camera configuration and in-place human motion only, losing track of human subjects' global motion trajectory. As elaborated in the previous paragraph, robust CVD is adopted to complete the missing information. Second, VIBE estimates body shapes for every video frame. This is unnecessary since people's body shapes are unlikely to change in a short period. Instead, we take the averaged body shapes for the same subject in a video sequence. 
	
	Finally, by combining the aforementioned steps, we can extract 3D body poses in a global frame and shape parameters from monocular RGB video, which can serve as input to generate SMPL body meshes.

	\subsection{From SMPL Models to IMU Data}
	Given the 3D human pose and shape represented by SMPL tri-mesh over time, accelerations and angular velocities in a global frame can be computed analytically. In particular, accelerations can be calculated by taking second derivatives of positions over time; angular velocities can be determined from the changes in the norm vector of a plane associated with three non-collinear mesh points (e.g., the vertices of a mesh triangle). However, SMPL tri-meshes generated by the models in Section~\ref{sc:video} tend to be noisy, erroneous and incomplete. Furthermore, accelerations and angular velocities measured by IMUs are subject to hardware imperfection such as noises, biases, and non-orthogonal axes, which are not easily replicated by analytical calculation. 
	
	To address the aforementioned issues, we design two neural network models, an accelerometer and a gyroscope network, to learn the mapping between motion trajectories of SMPL tri-mesh points and actual acceleration or angular velocity measured by IMUs in a global frame, respectively. The neural networks are capable of generating data from any arbitrary unseen region over the human body by training with real data from some selected on-body positions of various motion ranges (such as the head, chest, one side of the wrist, and ankle). Both models take the same design, with three convolutional and two bidirectional long-short term memory (LSTM) layers as the feature extractor, and a following linear layer as regression output. The model is fed a user-specified skin area, with three mesh triangles chosen near the area's center as input. In each triangle, the vertices are traversed counter-clockwise to ensure the norm direction always points outside of the human body. 
	
	\newcommand{\ba}[0]{\mathbf{a}}
	\newcommand{\bomega}[0]{\mathbf{\omega}}
	
	The collected IMU data are usually in the local sensor frame while the predictions of CROMOSim are in the global frame. Therefore, a coordinates transformation step is required. A user needs to select the skin region a virtual sensor affixes to and define its alignment represented as a rotation matrix ($R_S^B$). With the rotation matrix from the bone frame to the sensor frame $(R_S^B)^{-1}$, we can calculate IMU data in the sensor frame from the accelerations $\ba_G$ and angular velocities $\bomega_G$ in the global frame as follows:
	\begin{equation}\label{eq:gtos}
		\ba_S = (R_S^B)^{-1}\times(R_B^G)^{-1}\times(\ba_G+g),
	\end{equation}
	and 
	\begin{equation}
		\bomega_S = (R_S^B)^{-1}\times(R_B^G)^{-1}\times\bomega_G,
	\end{equation}
	where $R_B^G$ is obtained from the SMPL model for the corresponding skin region. 
	
	Due to noisy data sources and modeling errors, domain gaps exist between simulated and real data. Such gaps are more pronounced in simulated data from videos. To mitigate these gaps, we adopt the same distribution mapping technique \cite{conover1981rank} as IMUTube.
	
	\section{Evaluation}
	\label{section:Evaluation}
	In this section, we will evaluate CROMOSim in two sets of experiments. Firstly, we evaluate the fidelity of simulated sensor data both qualitatively and quantitatively. Then, we evaluate the utility of CROMOSim in data augmentation for downstream HAR tasks. 
	\subsection{Experimental Setup}
	\paragraph*{Datasets}
	\label{sc:totalcapture}
	To train the simulator network and evaluate the fidelity of simulated data, we use the TotalCapture dataset\cite{trumble2017total} which has all three data modalities (MoCap, IMU and video). For HAR evaluation, Realworld~\cite{realworld} and the Physical Activity Monitoring version 2 (PAMAP2) ~\cite{reiss2012introducing} datasets are used in task model training and testing. 
	
	In the fidelity evaluation, we divide data from all modalities into a 2-seconds sliding window with 80\% overlapping for model training and prediction. For HAR,  to make the results directly comparable to baseline approaches, we follow the same procedure described in IMUTube, where simulated and real IMU data are low-pass filtered, normalized and divided into sliding windows with 1-second length and 50\% overlapping. 
	
	\paragraph*{Evaluation metrics}
	To evaluate the fidelity of CROMOSim, we compute the root mean square error (RMSE) between simulated IMU data and ground truth. In HAR tasks, as the class in datasets is imbalanced, we adopt an F1 score to evaluate random single-subject-out experiments. 
	
	\paragraph*{Baseline Methods}
	We consider IMUSim and an analytic method as baselines to compare the fidelity of simulated data because IMUTube also utilizes IMUSim to generate IMU data from 3D global motion trajectories. The analytic method we adopt to compute linear acceleration is Richardson's extrapolation \cite{richardson1911approximate,richardson1927viii} . Compared to taking second-order derivatives, it gives a more accurate estimation with a 4th order error term (as opposed to 2nd order). The angular velocity of a selected skin region on an SMPL body mesh is calculated by tracking the rotation of its norm vector. 
	For HAR tasks, we take IMUTube as the baseline, but due to the lack of open source implementations, we include the reported performance on PAMAP2 dataset from \cite{kwon2020imutube}. 
	
	\subsection{Fidelity of CROMOSim\label{subsec:quality_mocap}}
	In this section, we provide qualitative and quantitative comparisons between CROMOSim and two baseline methods, namely, the analytical method (IMUCal) and IMUSim in terms of fidelity. We use TotalCapture in this experiment since it contains data from all three required modalities. Two CROMOSim models are trained using MoCap and video data from Subjects 1 -- 3  with sensor positions at their right wrist, right foot and pelvis. The models are used to predict accelerometer and gyroscope data on both left and right wrists of Subject 5 from the respective data sources. 
	
	\begin{figure}[h!]
		\begin{minipage}{\linewidth}
			\centering
			\includegraphics[width=0.49\linewidth]{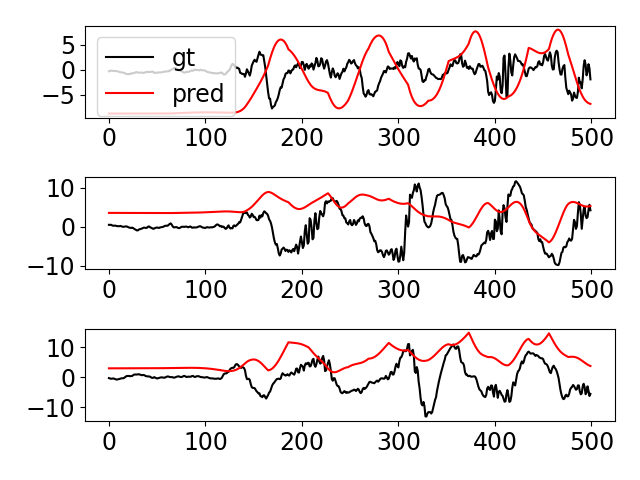}
			\includegraphics[width=0.49\linewidth]{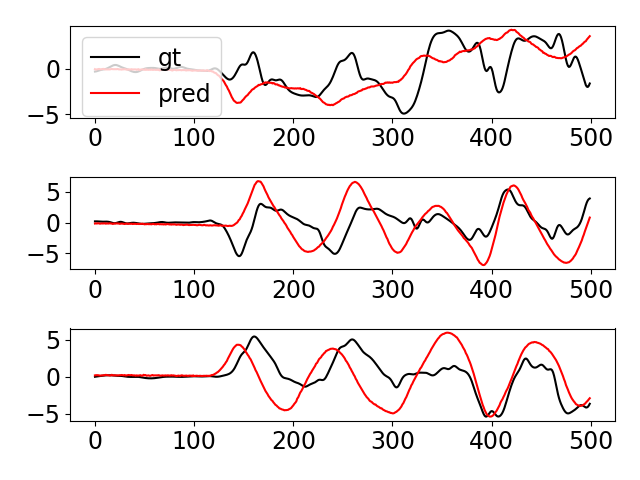}
			\subcaption{IMUSim}
			\includegraphics[width=0.49\linewidth]{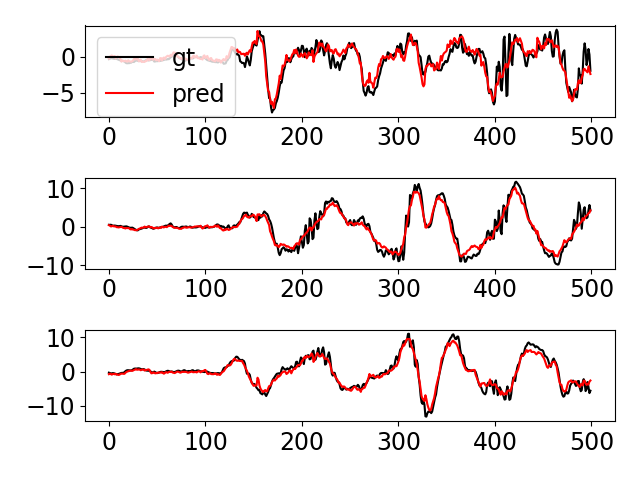}
			\includegraphics[width=0.49\linewidth]{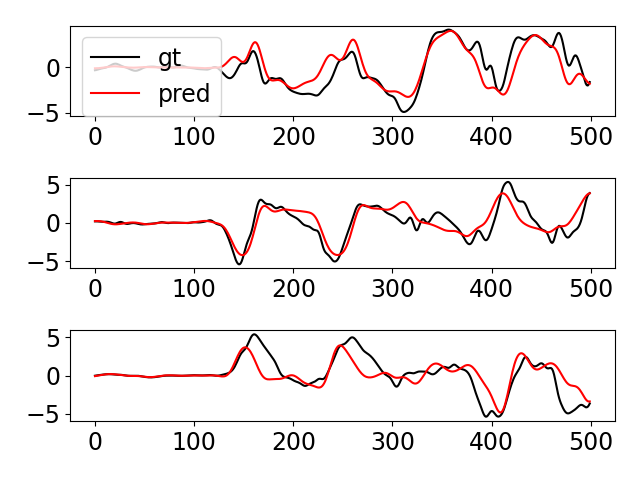}
			\subcaption{IMUCal}
			\includegraphics[width=0.49\linewidth]{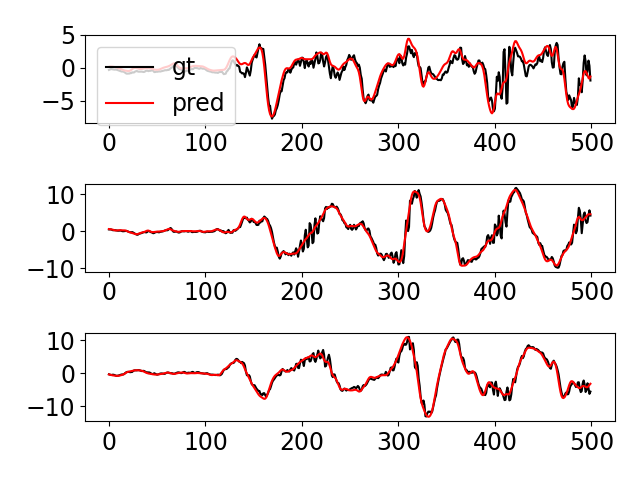}
			\includegraphics[width=0.49\linewidth]{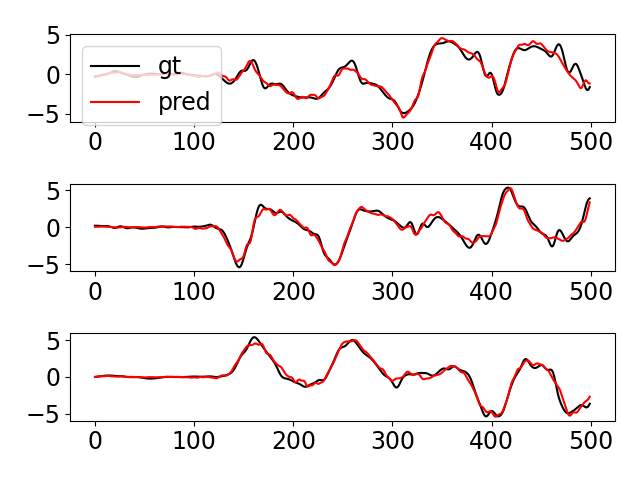}
			\subcaption{CROMOSim}	
		\end{minipage}
		
		\caption{Simulated IMU readings on the right wrist of Subject 5 from the MoCap data in TotalCapture. Left: accelerometer data. Right: gyroscope data.}
		\label{fg:mocap_plot}
	\end{figure}
	
	\begin{figure}[h!]
		\begin{minipage}{\linewidth}
			\centering
			\includegraphics[width=0.49\linewidth]{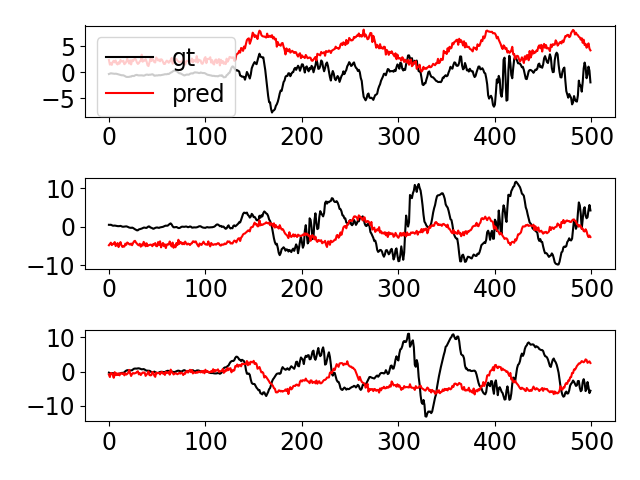}
			\includegraphics[width=0.49\linewidth]{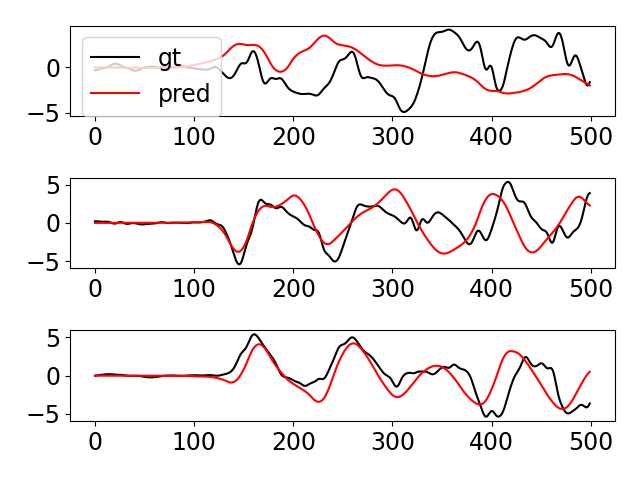}
			\subcaption{IMUSim}
			\includegraphics[width=0.49\linewidth]{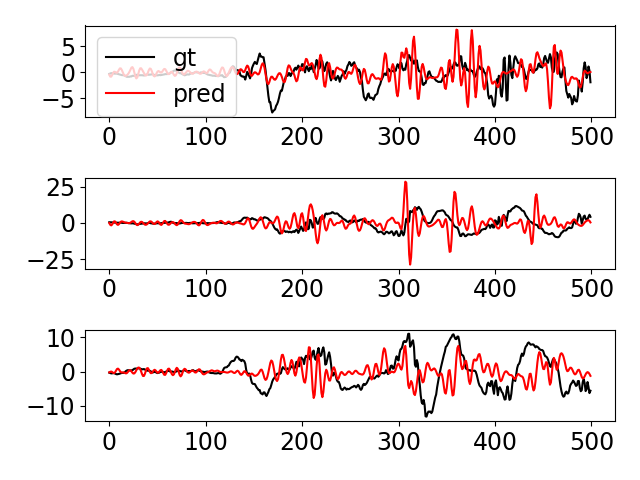}
			\includegraphics[width=0.49\linewidth]{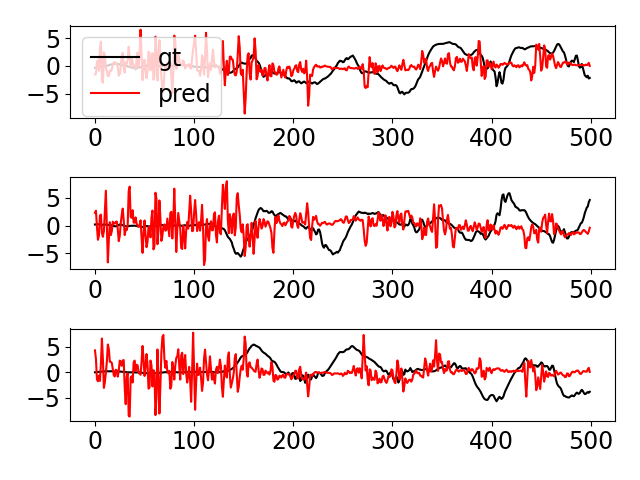}
			\subcaption{IMUCal}
			\includegraphics[width=0.49\linewidth]{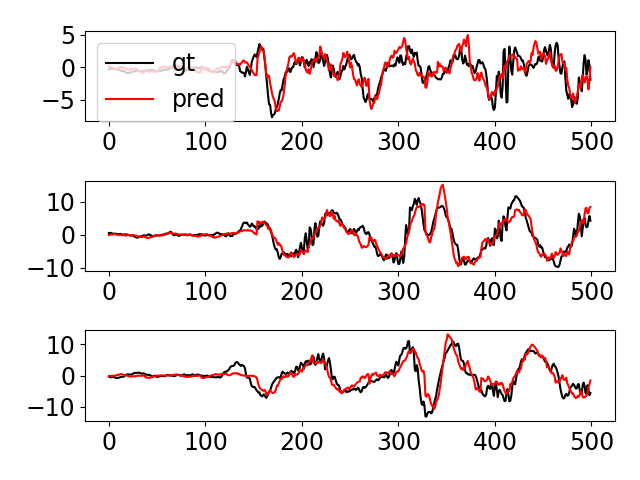}
			\includegraphics[width=0.49\linewidth]{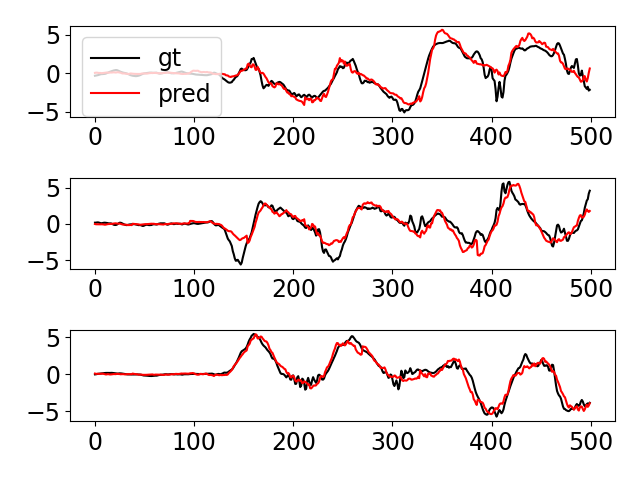}
			\subcaption{CROMOSim}
			
		\end{minipage}
		
		\caption{Simulated IMU readings on the right wrist of Subject 5 from monocular RGB camera video in TotalCapture. Left: accelerometer data. Right: gyroscope data.}
		\label{fg:video_plot}
	\end{figure}
	
	Figures~\ref{fg:mocap_plot} and \ref{fg:video_plot} show the simulated IMU readings from different methods with MoCap and RGB video data, respectively. From the figures, we observe that the fidelity of IMUSim is low across the board. It is because the default setting of IMUSim filters out too much high-frequency components. IMUCal works well for simulating accelerometer and gyroscope data with MoCap inputs. However, its performance significantly degrades when monocular RGB videos are taken as the source modality. This can be attributed to large noise and relative low accuracy of extracted SMPL body tri-mesh. In contrast, CROMOSim consistently outperforms baseline methods for both data modalities. 
	
	\begin{table}[t!]
		\centering
		\caption{\label{tb:rmse} RMSEs of simulated IMU readings on Subject 5's left wrist across all data trials.}
		\resizebox{\linewidth}{!}{%
			\begin{tabular}{l|l|l|l|l|l|l}
				\hline
				& \multicolumn{3}{c|}{Acceleration ($m/s^2$)}            & \multicolumn{3}{c}{Angular velocity ($rad/s$)}                \\ \hline 
				& IMUSim           & IMUCal & CROMOSim            & IMUSim           & IMUCal & CROMOSim            \\ \hline
				\begin{tabular}[c]{@{}l@{}}MoCap\\ extracted\\ SMPL\end{tabular} & 4.606 &1.785        & 1.602 &1.500   &1.272         & 0.801 \\ \hline
				\begin{tabular}[c]{@{}l@{}}Video\\ extracted\\ SMPL\end{tabular} &6.158  &  11.824      & 3.342 &1.848  &2.578        & 1.104 \\ \hline
			\end{tabular}
		}
	\end{table}	
	Table~\ref{tb:rmse} reports the case where both subject and sensor position are unseen to the simulator networks. The quantitative results are consistent with those in qualitative ones. With MoCap data, the accuracy of CROMOSim is 187.5\% and 11\% higher than that of IMUSim and IMUCal for accelerations, respectively,  and 87\% and 58\% for gyroscope data. The advantage of CROMOSim is more pronounced with monocular RGB videos, outperforming the next best method (IMUSim) by 84\% and 67\% for accelerometer and gyroscope data. 
	
	\subsection{Applications of CROMOSim in HAR Tasks}
	\label{sect:downstream tasks}
	In this section, we evaluate the utility of CROMOSim in data augmentation for training HAR models. Here we consider three settings: i) R2R, where models are both trained and tested with real IMU data; ii) V2R, where models are trained with simulated data but tested with real data; iii) Mix2R, where models are trained using a mixture of real and simulated data while tested with real data.
	
	We adopt the DeepConvLSTM network proposed in ~\cite{ordonez2016deepconvlstm} as the task model, while the same simulator neural network trained on the TotalCapture dataset is used here to simulate sensor readings from videos. Evaluations are made on the Realworld and PAMAP2 datasets respectively, with data simulated from the same video source (Realworld dataset). An ablation study was conducted by removing robust CVD from the proposed pipeline, and the resulting approach is called {\it CROMOSim Lite.} To make the result directly comparable, we followed the experiment protocol in IMUTube \cite{kwon2020imutube}. 
	
	\begin{table}[t!]
		\centering
		\caption{\label{tb:realworld}Average F1 scores of random single-subject-hold out experiments on the RealWorld dataset. IMUTube$^\star$ corresponds the scores reported in \cite{kwon2020imutube}}
		\begin{tabular}{l|lll}
			\hline
			&	R2R & V2R & Mix2R \\ \hline
			IMUTube$^\star$&	0.730$\pm$0.007 & 0.546$\pm$0.008 & 0.778$\pm$0.007  \\ \hline
			IMUTube&	0.729$\pm$0.007 & 0.552$\pm$0.005 & 0.781$\pm$0.011  \\ \hline
			CROMOSim Lite&	 0.729$\pm$0.007 & { 0.580$\pm$0.047} & { 0.802 $\pm$0.013}  \\ \hline
			CROMOSim&	 0.729$\pm$0.007 & {\bf 0.593$\pm$0.012} & {\bf  0.821$\pm$0.003}  \\ \hline
		\end{tabular}
	\end{table}
	Table~\ref{tb:realworld} reports the average F1 scores of five single-subject-hold out experiments on the RealWorld dataset. Since the authors of IMUTube provide their simulated data on this dataset, we directly replicated their experiments and the results are in the second row. For comparison, we also include the scores reported in \cite{kwon2020imutube} as the first row. It can be seen the two are quite similar to one another. Even CROMOSim Lite outperforms IMUTube in V2R and Mix2R experiments, while CROMOSim works the best. Moreover, Mix2R achieves much higher F1 scores compared to R2R and V2R, demonstrating the utility of data augmentation with simulated data.  
	
	\begin{table}[t!]
		\centering
		\caption{\label{tb:pamap2}Random single subject hold out evaluation on PAMAP2 dataset with mean F1-score. IMUTube$^\star$ corresponds to the scores reported in \cite{kwon2020imutube}}
		\begin{tabular}{l|lll}
			\hline
			&	R2R & V2R & Mix2R \\ \hline
			IMUTube$^\star$&	0.700$\pm$0.016 &0.552$\pm$0.017  & 0.702$\pm$0.016  \\ \hline
			CROMOSim Lite&	 0.702$\pm$0.021 & { 0.638$\pm$0.009} & { 0.726$\pm$0.014}  \\ \hline
			CROMOSim&	0.702$\pm$0.021 & {\bf 0.689$\pm$0.012} & {\bf 0.769$\pm$0.009}  \\ \hline
		\end{tabular}
	\end{table}
	
	Table~\ref{tb:pamap2} summarizes the results from CROMOSim and those reported in \cite{kwon2020imutube}. Due to the different sensor placements in the PAMAP2 datasets, the simulated data provided by the authors of IMUTube cannot be used. Similar to the RealWorld dataset, CROMOSim outperforms IMUTube for the PAMAP2 datasets but with a more prominent margin; the HAR model trained from Mix2R is still superior to those from R2R and V2R. 
	
	\section{Conclusion and Future Work}
	\label{section:Conclusion}
	In this paper, we implemented CROMOSim, a pipeline that simulates accelerometer and gyroscope readings at arbitrary user-designated on-body positions from MoCap and monocular RGB camera videos. A DNN model is trained to learn the functional mapping between imperfect trajectory estimations in a 3D body tri-mesh to IMU data. Experiments showed that CROMOSim can generate higher fidelity data than baseline methods and is useful for downstream HAR tasks. As part of the future work, we are implementing a graphical user interface and wrapping up CROMOSim as an easy-to-use tool now. Hopefully, it will be open-sourced to the public by this summer.
	
	\bibliographystyle{IEEEtran}
	\bibliography{ref}

\begin{thebibliography}{10}
\providecommand{\url}[1]{#1}
\csname url@samestyle\endcsname
\providecommand{\newblock}{\relax}
\providecommand{\bibinfo}[2]{#2}
\providecommand{\BIBentrySTDinterwordspacing}{\spaceskip=0pt\relax}
\providecommand{\BIBentryALTinterwordstretchfactor}{4}
\providecommand{\BIBentryALTinterwordspacing}{\spaceskip=\fontdimen2\font plus
\BIBentryALTinterwordstretchfactor\fontdimen3\font minus
  \fontdimen4\font\relax}
\providecommand{\BIBforeignlanguage}[2]{{%
\expandafter\ifx\csname l@#1\endcsname\relax
\typeout{** WARNING: IEEEtran.bst: No hyphenation pattern has been}%
\typeout{** loaded for the language `#1'. Using the pattern for}%
\typeout{** the default language instead.}%
\else
\language=\csname l@#1\endcsname
\fi
#2}}
\providecommand{\BIBdecl}{\relax}
\BIBdecl

\bibitem{wang2019deep}
J.~Wang, Y.~Chen, S.~Hao, X.~Peng, and L.~Hu, ``Deep learning for sensor-based
  activity recognition: A survey,'' \emph{Pattern Recognition Letters}, vol.
  119, pp. 3--11, 2019.

\bibitem{gu2021survey}
F.~Gu, M.-H. Chung, M.~Chignell, S.~Valaee, B.~Zhou, and X.~Liu, ``A survey on
  deep learning for human activity recognition,'' \emph{ACM Computing Surveys
  (CSUR)}, vol.~54, no.~8, pp. 1--34, 2021.

\bibitem{vaizman2017recognizing}
Y.~Vaizman, K.~Ellis, and G.~Lanckriet, ``Recognizing detailed human context in
  the wild from smartphones and smartwatches,'' \emph{IEEE pervasive
  computing}, vol.~16, no.~4, pp. 62--74, 2017.

\bibitem{reiss2012introducing}
A.~Reiss and D.~Stricker, ``Introducing a new benchmarked dataset for activity
  monitoring,'' in \emph{2012 16th International Symposium on Wearable
  Computers}.\hskip 1em plus 0.5em minus 0.4em\relax IEEE, 2012, pp. 108--109.

\bibitem{mahmood2019amass}
N.~Mahmood, N.~Ghorbani, N.~F. Troje, G.~Pons-Moll, and M.~J. Black, ``Amass:
  Archive of motion capture as surface shapes,'' in \emph{Proceedings of the
  IEEE/CVF International Conference on Computer Vision}, 2019, pp. 5442--5451.

\bibitem{loper2015smpl}
M.~Loper, N.~Mahmood, J.~Romero, G.~Pons-Moll, and M.~J. Black, ``Smpl: A
  skinned multi-person linear model,'' \emph{ACM transactions on graphics
  (TOG)}, vol.~34, no.~6, pp. 1--16, 2015.

\bibitem{young2011imusim}
A.~D. Young, M.~J. Ling, and D.~K. Arvind, ``Imusim: A simulation environment
  for inertial sensing algorithm design and evaluation,'' in \emph{Proceedings
  of the 10th ACM/IEEE International Conference on Information Processing in
  Sensor Networks}.\hskip 1em plus 0.5em minus 0.4em\relax IEEE, 2011, pp.
  199--210.

\bibitem{huang2018deep}
Y.~Huang, M.~Kaufmann, E.~Aksan, M.~J. Black, O.~Hilliges, and G.~Pons-Moll,
  ``Deep inertial poser: Learning to reconstruct human pose from sparse
  inertial measurements in real time,'' \emph{ACM Transactions on Graphics
  (TOG)}, vol.~37, no.~6, pp. 1--15, 2018.

\bibitem{xiao2020deep}
F.~Xiao, L.~Pei, L.~Chu, D.~Zou, W.~Yu, Y.~Zhu, and T.~Li, ``A deep learning
  method for complex human activity recognition using virtual wearable
  sensors,'' in \emph{International Conference on Spatial Data and
  Intelligence}.\hskip 1em plus 0.5em minus 0.4em\relax Springer, 2020, pp.
  261--270.

\bibitem{takeda2018multi}
S.~Takeda, T.~Okita, P.~Lago, and S.~Inoue, ``A multi-sensor setting activity
  recognition simulation tool,'' in \emph{Proceedings of the 2018 ACM
  International Joint Conference and 2018 International Symposium on Pervasive
  and Ubiquitous Computing and Wearable Computers}, 2018, pp. 1444--1448.

\bibitem{liu2021video}
Y.~Liu, S.~Zhang, and M.~Gowda, ``When video meets inertial sensors: Zero-shot
  domain adaptation for finger motion analytics with inertial sensors,'' in
  \emph{Proceedings of the International Conference on Internet-of-Things
  Design and Implementation}, 2021, pp. 182--194.

\bibitem{rey2019let}
V.~F. Rey, P.~Hevesi, O.~Kovalenko, and P.~Lukowicz, ``Let there be imu data:
  generating training data for wearable, motion sensor based activity
  recognition from monocular rgb videos,'' in \emph{Adjunct Proceedings of the
  2019 ACM International Joint Conference on Pervasive and Ubiquitous Computing
  and Proceedings of the 2019 ACM International Symposium on Wearable
  Computers}, 2019, pp. 699--708.

\bibitem{rey2020yet}
V.~F. Rey, K.~K. Garewal, and P.~Lukowicz, ``Yet it moves: Learning from
  generic motions to generate imu data from youtube videos,'' \emph{arXiv
  preprint arXiv:2011.11600}, 2020.

\bibitem{kwon2020imutube}
H.~Kwon, C.~Tong, H.~Haresamudram, Y.~Gao, G.~D. Abowd, N.~D. Lane, and
  T.~Ploetz, ``Imutube: Automatic extraction of virtual on-body accelerometry
  from video for human activity recognition,'' \emph{Proceedings of the ACM on
  Interactive, Mobile, Wearable and Ubiquitous Technologies}, vol.~4, no.~3,
  pp. 1--29, 2020.

\bibitem{kwon2021approaching}
H.~Kwon, B.~Wang, G.~D. Abowd, and T.~Pl{\"o}tz, ``Approaching the real-world:
  Supporting activity recognition training with virtual imu data,''
  \emph{Proceedings of the ACM on Interactive, Mobile, Wearable and Ubiquitous
  Technologies}, vol.~5, no.~3, pp. 1--32, 2021.

\bibitem{kopf2021rcvd}
J.~Kopf, X.~Rong, and J.-B. Huang, ``Robust consistent video depth
  estimation,'' in \emph{IEEE/CVF Conference on Computer Vision and Pattern
  Recognition}, 2021.

\bibitem{mun2019unsupervised}
J.-H. Mun, M.~Jeon, and B.-G. Lee, ``Unsupervised learning for depth,
  ego-motion, and optical flow estimation using coupled consistency
  conditions,'' \emph{Sensors}, vol.~19, no.~11, p. 2459, 2019.

\bibitem{luo2020consistent}
X.~Luo, J.-B. Huang, R.~Szeliski, K.~Matzen, and J.~Kopf, ``Consistent video
  depth estimation,'' \emph{ACM Transactions on Graphics (TOG)}, vol.~39,
  no.~4, pp. 71--1, 2020.

\bibitem{prince2012computer}
S.~J. Prince, \emph{Computer vision: models, learning, and inference}.\hskip
  1em plus 0.5em minus 0.4em\relax Cambridge University Press, 2012.

\bibitem{cao2019openpose}
Z.~Cao, G.~Hidalgo, T.~Simon, S.-E. Wei, and Y.~Sheikh, ``Openpose: realtime
  multi-person 2d pose estimation using part affinity fields,'' \emph{IEEE
  transactions on pattern analysis and machine intelligence}, vol.~43, no.~1,
  pp. 172--186, 2019.

\bibitem{kocabas2020vibe}
M.~Kocabas, N.~Athanasiou, and M.~J. Black, ``Vibe: Video inference for human
  body pose and shape estimation,'' in \emph{Proceedings of the IEEE/CVF
  Conference on Computer Vision and Pattern Recognition}, 2020, pp. 5253--5263.

\bibitem{conover1981rank}
W.~J. Conover and R.~L. Iman, ``Rank transformations as a bridge between
  parametric and nonparametric statistics,'' \emph{The American Statistician},
  vol.~35, no.~3, pp. 124--129, 1981.

\bibitem{trumble2017total}
M.~Trumble, A.~Gilbert, C.~Malleson, A.~Hilton, and J.~P. Collomosse, ``Total
  capture: 3d human pose estimation fusing video and inertial sensors.'' in
  \emph{BMVC}, vol.~2, no.~5, 2017, pp. 1--13.

\bibitem{realworld}
T.~Sztyler and H.~Stuckenschmidt, ``On-body localization of wearable devices:
  An investigation of position-aware activity recognition,'' in \emph{2016 IEEE
  International Conference on Pervasive Computing and Communications (PerCom)},
  2016, pp. 1--9.

\bibitem{richardson1911approximate}
L.~F. Richardson, ``The approximate arithmetical solution by finite differences
  with an application to stresses in masonry dams,'' \emph{Philosophical
  Transactions of the Royal Society of America}, vol. 210, pp. 307--357, 1911.

\bibitem{richardson1927viii}
L.~F. Richardson and J.~A. Gaunt, ``Viii. the deferred approach to the limit,''
  \emph{Philosophical Transactions of the Royal Society of London. Series A,
  containing papers of a mathematical or physical character}, vol. 226, no.
  636-646, pp. 299--361, 1927.

\bibitem{ordonez2016deepconvlstm}
F.~Ord{\'o}{\~n}ez and D.~Roggen, ``Deep convolutional and lstm recurrent
  neural networks for multimodal wearable activity recognition,''
  \emph{Sensors}, vol.~16, no.~1, p. 115, 2016.

\end{thebibliography}
	
\end{document}